\documentclass[twocolumn, superscriptaddress, amsmath, secnumarabic, floatfix, amssymb, nobibnotes, aps, pra, reprint]{revtex4-1}

\usepackage{microtype}
\usepackage{graphicx}
\usepackage{xcolor}
\usepackage[utf8]{inputenc}
\begin{document}

\title{A direct GHz-clocked phase and intensity modulated transmitter applied to quantum key distribution}

\author{G. L. Roberts}%
\email[Corresponding author: ]{zhiliang.yuan@crl.toshiba.co.uk}
\affiliation{Toshiba Research Europe Ltd, 208 Cambridge Science Park, Milton Road, Cambridge, CB4 0GZ, United Kingdom}
\affiliation{Cambridge University Engineering Department, 9 JJ Thomson Avenue, Cambridge, CB3 0FA, United Kingdom}
\author{M. Lucamarini}%
\affiliation{Toshiba Research Europe Ltd, 208 Cambridge Science Park, Milton Road, Cambridge, CB4 0GZ, United Kingdom}
\author{J. F. Dynes}%
\affiliation{Toshiba Research Europe Ltd, 208 Cambridge Science Park, Milton Road, Cambridge, CB4 0GZ, United Kingdom}
\author{S. J. Savory}%
\affiliation{Cambridge University Engineering Department, 9 JJ Thomson Avenue, Cambridge, CB3 0FA, United Kingdom}
\author{Z. L. Yuan}%
\affiliation{Toshiba Research Europe Ltd, 208 Cambridge Science Park, Milton Road, Cambridge, CB4 0GZ, United Kingdom}
\author{A. J. Shields}%
\affiliation{Toshiba Research Europe Ltd, 208 Cambridge Science Park, Milton Road, Cambridge, CB4 0GZ, United Kingdom}

\begin{abstract}
Quantum key distribution (QKD), a technology that enables perfectly secure communication, has evolved to the stage where many different protocols are being used in real-world implementations.
Each protocol has its own advantages, meaning that users can choose the one best-suited to their application, however each often requires different hardware.
This complicates multi-user networks, in which users may need multiple transmitters to communicate with one another.
Here, we demonstrate a direct-modulation based transmitter that can be used to implement most weak coherent pulse based QKD protocols with simple changes to the driving signals.
This also has the potential to extend to classical communications, providing a low chirp transmitter with simple driving requirements that combines phase shift keying with amplitude shift keying.
We perform QKD with concurrent time-bin and phase modulation, alongside phase randomisation.
The acquired data is used to evaluate secure key rates for time-bin encoded BB84 with decoy states and a finite key-size analysis, giving megabit per second secure key rates, 1.60 times higher than if purely phase-encoded BB84 was used.
\end{abstract}
\maketitle

\section{Introduction}

Quantum key distribution (QKD) allows users to communicate with information theoretic security~\cite{Gisin_quantum_2002, Scarani_security_2009}.
This is possible by encoding the key on single photons so that a malign party trying to measure a key bit will alter its state in a manner observable to the legitimate parties.
The security provided is of great value to anyone wishing to future-proof the secrecy of their information transfer.
The technology is also practical and is currently implemented in a number of metropolitan networks~\cite{Peev_SECOQC_2009, Sasaki_Field_2011} and even in ground-satellite links~\cite{Vallone_Experimental_2015, Liao_Long-distance_2017, Takenaka_satellite--ground_2017}.

Research developments tend to aim at improving the secure key rate and the achievable distance of QKD systems~\cite{Ma_practical_2005, Dixon_Gigahertz_2008, Bacco_two-dimensional_2016, Yin_measurement-device-independent_2016}.
For example, the decoy-state BB84 protocol has security against coherent attacks, is able to reach distances of hundreds of kilometres and can achieve megabit per second secure key rates~\cite{Lim_concise_2014, Frohlich_long-distance_2017}.
However, this often makes systems more complex, requiring stabilisation routines and extra consideration to protect against side channels, where Eve attacks the practical implementation~\cite{Dixon_quantum_2017, Lo_Measurement-device-independent_2012}.

\begin{figure*}[htbp]
\centering
\includegraphics[width=\linewidth]{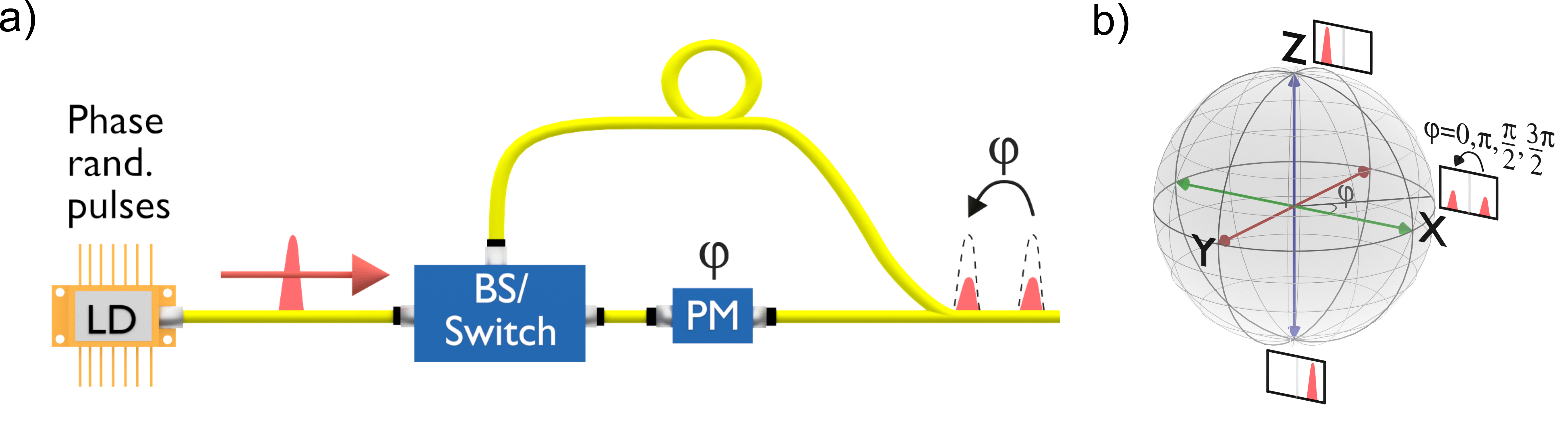}
\caption{\textbf{X, Y, Z QKD bases} a) A possible schematic to produce all the necessary QKD states using a phase-randomised pulse source and a device that can act as a beamsplitter (BS) and a high-speed switch, plus a phase modulator (PM) producing a phase shift $\phi$.
To produce the X and Y bases, the device acts as a beamsplitter (BS), phase modulating one half of the pulse using the PM and delaying the other half with an optical delay line.
To produce the Z basis, the device acts as a switch, routing the pulse down one path to place it in the desired time bin and b) Bloch sphere representation of qubit states. }
\label{fig:ZXY}
\end{figure*}

QKD can be carried out using orthogonal states within two or more non-orthogonal bases.
This means the result is non-deterministic if the state encoded in a certain basis is measured in a different basis.
Time-bin qubits, prepared with the setup in Figure~\ref{fig:ZXY}a, are the natural choice in optical fibres because the pulses travel along the fibre with their phase reference, meaning that perturbations apply to both pulses.
Their state can be conveniently represented using the Bloch sphere, as depicted in Figure~\ref{fig:ZXY}b.
The equatorial bases, X and Y, correspond to two equal intensity pulses with a phase difference between them.
States in X and Y can be realised by separating a single phase-randomised pulse into a signal and reference pulse using an asymmetric Mach-Zehnder interferometer (AMZI), then encoding a phase difference using a phase modulator.
This can be decoded using an identical AMZI.
The polar basis, Z, corresponds to a pulse in just one of the two potential time bins.
States in this basis can be decoded by measuring the arrival time of the time-bins in the receiver's detectors.

In this manuscript, we refer to the QKD protocol using the Z basis alongside either the X or Y bases as \textit{polar BB84}, and the protocol using the X and Y bases as \textit{phase-encoded BB84}.
Polar BB84 could theoretically be implemented using the BS/switch component shown in Figure~\ref{fig:ZXY}a.
However this is not a commonly available component and it is challenging to build a device that can act reliably as a high-speed switch and beamsplitter at the rates required by QKD systems.
One of the most practical setups to implement polar BB84~\cite{Yoshino_quantum_2018} uses a phase-randomised pulsed laser diode as the source, separated into a reference and signal pulse by an AMZI and then encoded using another intensity modulator (IM) and a phase modulator (PM).
This setup is bulky and would require stabilisation routines to ensure the AMZI delay line is matched to that in Bob's receiver for a real-world implementation.
Another drawback is that the transmitter is not versatile, requiring modifications if one wishes to implement another QKD protocol, for example differential phase shift~\cite{Shibata_quantum_2014, Inoue_differential_2015}, coherent one way~\cite{Korzh_provably_2015, Branciard_upper_2008} or differential quadrature phase shift~\cite{Inoue_differential-quadrature-phase-shift_2009, Kawakami_security_2016}.

A promising QKD transmitter that mitigates these aforementioned drawbacks modulates the phase of one laser, which is then inherited by the pulses of another laser via optical injection locking (OIL)~\cite{Yuan_directly_2016}.
This enables the precise control of the output phase of pulses, as well as the ability to perform on-demand phase randomisation~\cite{Roberts_manipulating_2017}.
OIL also gives an enhanced modulation bandwidth, allowing time-bin encoding to be carried out on gain-switched pulses at 2~GHz, whilst maintaining a coherent phase~\cite{Roberts_modulator-free_2017}.
However, it has not yet been possible to simultaneously directly modulate the phase and intensity of a light source with the high purity necessary for QKD.

Here, we use direct laser modulation to concurrently modulate the phase and intensity of the transmitter to provide six states that can be used to perform QKD without the need for an interferometer in the transmitter.
The directly-modulated system produces signal and vacuum states, allowing a single IM to be used to prepare the decoy states and to equalise the mean photon number in the phase bases.
We use the Z and X bases to implement the polar BB84 protocol and compare the results to phase-encoded BB84 implemented with the Y and X bases.
The low quantum bit error rate (QBER) of the polar basis relative to the equatorial bases means that its use as the signal state allows for fewer bits to be lost to error correction, enhancing the secure key rate.

\section{Experimental Realization}

\begin{figure}[hb]
\centering
\includegraphics[width=\linewidth]{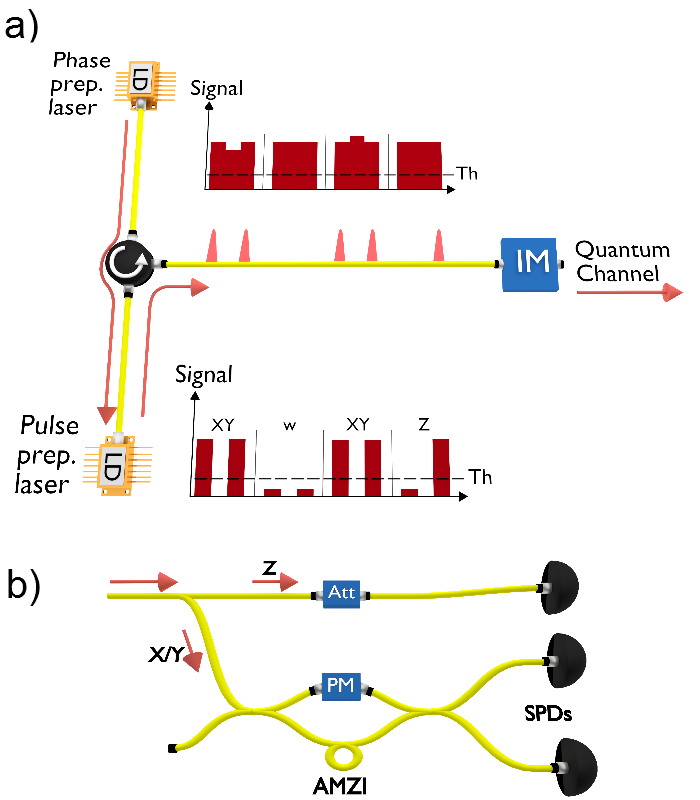}
\caption{\textbf{Experimental design.}
a) Light from the phase preparation laser diode (LD) is injected into the pulse preparation laser diode via a circulator.
The applied electrical signals are shown beside each laser and the pulse intensities are shown along the fibre.
An intensity modulator (IM) can then be used to prepare decoy states and to equalise the polar and equatorial state intensities before attenuation and transmission through the quantum channel.
\textcolor{black}{b) The Z basis is measured through direct detection whereas X/Y basis detection uses an asymmetric Mach-Zehnder interferometer (AMZI) with a phase modulator (PM) on one arm. An attenuator (Att) on the Z detection arm balances the losses of the X/Y detection arm. }}
\label{fig:experimentalDesign}
\end{figure}

The transmitter we use is based on OIL and is shown in Figure~\ref{fig:experimentalDesign}.
The protocols implemented are decoy-state polar BB84 and decoy-state phase-encoded BB84~\cite{Hwang_quantum_2003, Wang_beating_2005, Lo_decoy_2005}.

The phase preparation laser encodes a relative phase between pulse pairs using a 750~ps signal to bring the laser above threshold and to coincide temporally with two pulse preparation laser pulses 500~ps apart.
A 250~ps modulation is applied in the middle of this signal to control the relative phase between the two pulses.
The laser is then driven below threshold for 250~ps to ensure the global phase of every pulse pair is completely random due to the random phase of spontaneous emission photons~\cite{Yuan_directly_2016, Roberts_manipulating_2017}.
The optical signal is injected into the pulse preparation laser via a circulator, where pulses adopt the phase of the phase preparation laser.
This removes the need for a high speed phase modulator and an extra random number generator for phase-encoding and randomisation.
\textcolor{black}{A 1550~nm DFB laser diode with a 10~GHz bandwidth (Gooch \& Housego AA0701) is used as the phase preparation laser and a custom-made laser without an optical isolator as the pulse preparation laser.}

The electrical signal into the pulse preparation laser is patterned to produce intensity-modulated gain-switched pulses~\cite{Roberts_modulator-free_2017}.
For polar BB84 with decoy states, empty pulses are required to prepare Z basis states and vacuum states.
250~ps electrical pulses are input to the pulse preparation laser at a frequency of 2~GHz and the DC is set to below the lasing threshold.
When a signal or decoy pulse is required, the electrical signal is above the lasing threshold.
To prepare a vacuum state the electrical signal is below the lasing threshold, as shown in Figure~\ref{fig:experimentalDesign}.
The X and Y bases can then be attenuated by 3~dB so that they contain the same mean photon number as the Z basis.
Although only two bases are used for BB84, we take experimental data for the X, Y and Z bases, allowing us to demonstrate the versatility of the transmitter.

The transmission basis probabilities are set to $P_Z = 0.8$, $P_X = P_Y = 0.1$ and the probabilities of sending a signal (photon flux s), decoy (photon flux v) and vacuum (photon flux w) state are $P_s = 0.8$, $P_v = P_w = 0.1$ respectively.
The photon fluxes are 0.5, 0.038 and 0.001 for s, v and w respectively.
A  proof of principle experiment is then carried out, where data is measured for 20 minutes per basis at each distance, giving 40 minutes of key time for both the polar BB84 protocol and phase-encoded BB84.
This allows us to maximise the number of key bits, whilst providing a sufficient number of bits in the check basis to keep the fluctuations low.

The pulse preparation laser is clocked at 2~GHz, giving an effective system clock rate of 1~GHz.
This is because two time bins are required to encode a single qubit.
A 2$^{10}$-bit pseudorandom sequence is generated as Alice's pattern, allowing the corresponding electrical signals to be input to drive the laser diodes.
A fixed 12~GHz spectral filter \textcolor{black}{(Advanced Optics Solutions -- ASE Filter) at 1550.12~nm} is placed at Alice's output to reduce any amplified spontaneous emission.
The pulses are then attenuated to the required photon number before being sent through the quantum channel to Bob.

\begin{figure}[ht]
\centering
\includegraphics[width=\linewidth]{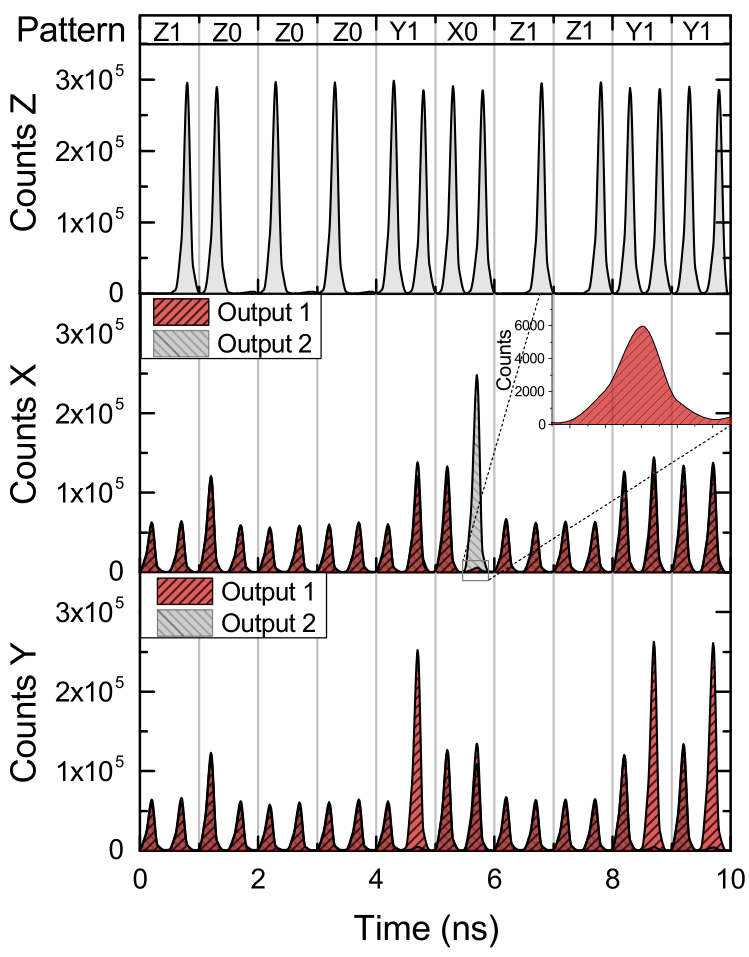}
\caption{\textbf{Directly-modulated traces}. Measurement traces before decoy state preparation and basis intensity equalisation. The Z basis (top) has only a single trace, whereas the X and Y bases (middle and bottom) have two AMZI outputs. The corresponding input pattern values are displayed at the top, labelled as Bb, where B is the basis and b is the logical bit value inside that basis. A red (grey) peak in the X and Y bases corresponds to a `0' (`1') logical bit, where the photon exits through the upper (lower) AMZI port. Peaks in output 1 (2) counts of the AMZI are complemented by small counts in output 2 (1) (middle inset), showing the high distinguishability between bits. }
\label{fig:traces}
\end{figure}

The X and Y data are collected using an AMZI with a 500~ps time delay on one arm to interfere consecutive pulses.
A polariser is placed at the output of the AMZI to clean the signal, necessitating the use of a polarisation controller in Alice for alignment.
The AMZI has a 1.7~dB loss and half of the photons (the reference pulses) contain no information so are discarded.
A fixed attenuation of 4.7~dB must be placed on the Z measurement arm to balance the detection efficiencies for each measurement basis.
\textcolor{black}{This is because the security of BB84 relies on identical basis-independent detection probabilities~\cite{Koashi_efficient_2006, Scarani_security_2009, Tomamichel_tight_2012}. }
The detected counts are tagged using a digitizer and binned into 2$^{11}$-bin histograms for extraction of the counts and error rates.
A 10~ns subset of this histogram before creation of decoy states and equalisation of the X and Y bases intensities can be seen in Figure~\ref{fig:traces}.
Random interference occurs in the X and Y bases when two pulses from separate blocks interfere, giving an average interference intensity of half the maximum intensity.
An empty pulse followed by a full pulse has no interference, thus producing a quarter of the maximum intensity.

The detectors used are superconducting nanowire SPDs with a detection efficiency of 34~\%, a dark count rate of 30~Hz \textcolor{black}{and a deadtime of $<$20~ns.
The jitter increases and efficiency decreases with increasing count rate, so measurements are not taken at count rates over 10~MCounts/s. }
A digitizer with 100~ps time bins and constant factor discrimination then processes the counts.
The detectors have a polarisation dependence, so a polarisation controller is used for optimal detection efficiency.
Although the receiver is adapted to each specific protocol, the transmitter remains entirely unchanged.
This is a necessary feature for a multi-protocol QKD transmitter.

\section{Results}

\begin{figure}[ht]
\centering
\includegraphics[width=\linewidth]{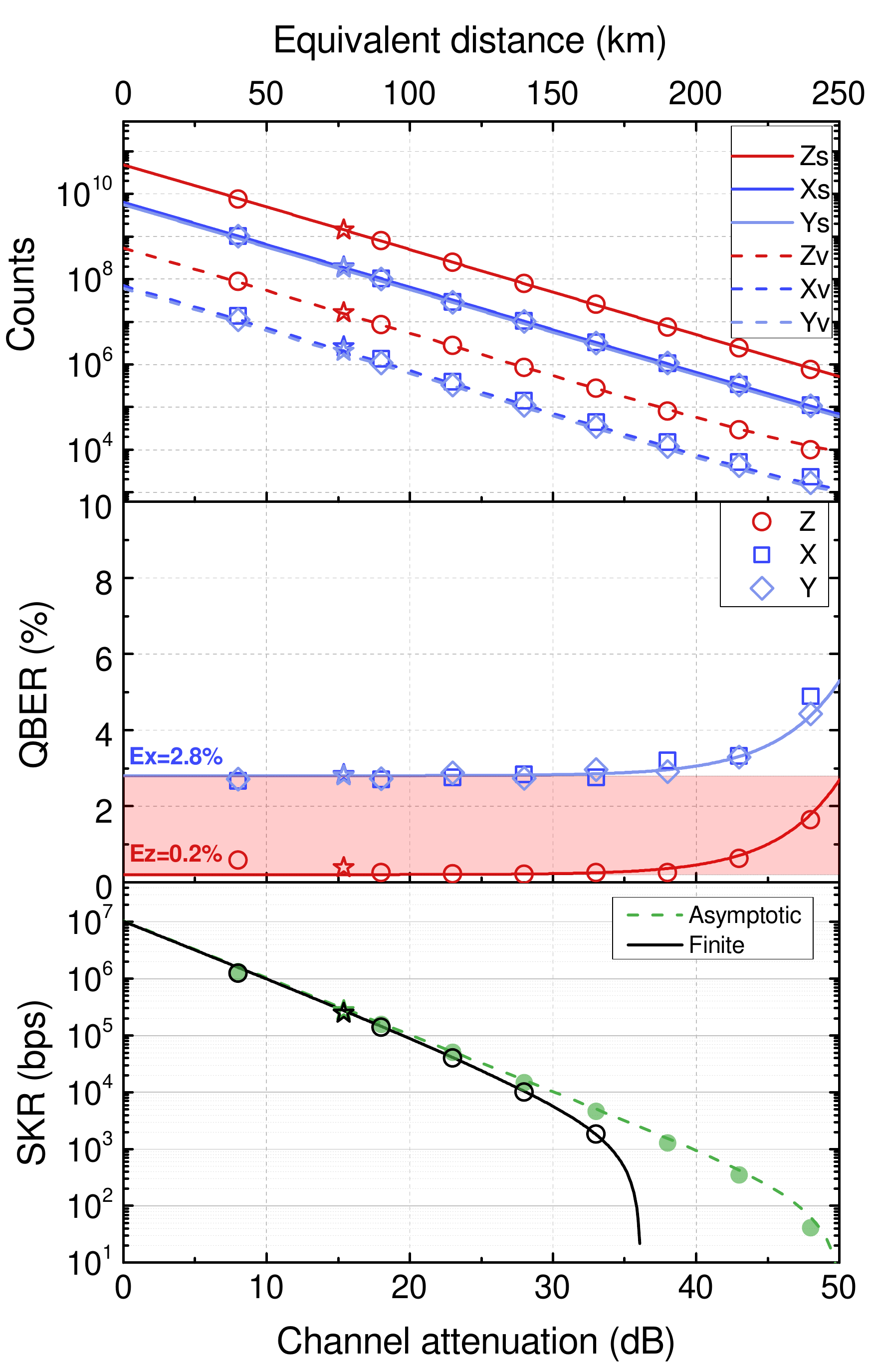}
\caption{\textbf{Polar BB84 results} with decoy states, secure against coherent attacks. Lines are theoretical results and symbols are experimentally measured or calculated values. a) Number of counts measured in each basis, b) QBER for each basis, c) Secure key rate (SKR) in the asymptotic (filled symbols, dotted line) and finite key-size regimes (empty symbols, solid line). The star symbols are data measured with real fibre as the quantum channel. }
\label{fig:keyRates}
\end{figure}

The number of signal and decoy counts measured in each basis is shown in Figure~\ref{fig:keyRates}a).
These decrease exponentially because the measurement time remains constant at 20 minutes, regardless of the distance, whereas received counts scales exponentially with channel loss.
Figure~\ref{fig:keyRates}b) highlights the 2.6 percentage point drop in QBER between the XY bases and Z basis.
Simulations using the experimental parameters and the predicted count rate based on the system losses are also shown.
The finite key-size analysis is detailed by Lim et al~\cite{Lim_concise_2014}, which quantifies the security and correctness of the protocol through the parameters $\varepsilon_{\textrm{sec}}$ and $\varepsilon_{\textrm{cor}}$.
In this implementation, these values are set to $2 \times 10^{-11}$ and $1 \times 10^{-15}$ respectively.
The key rate, R$_L$ is calculated using
\begin{equation}
\label{eq:keyRate}
R_{L} = \left[
s_{ZZ; 0} + s_{ZZ; 1} (1 - \textrm{h}(\phi_z)) - \lambda_{EC} -
\Delta(\varepsilon_{\textrm{sec}}, \varepsilon_{\textrm{cor}})
\right]/t,
\end{equation}
where $s_{XX, ZZ; n}$ is the number of counts measured by Bob in the X or Z basis, given that Alice prepared an \textit{n}-photon state in the X or Z basis respectively, $\phi_Z$ is the single photon phase error rate in Z, $\lambda_{EC}$ is the error correction information, $\Delta$ is the finite key-size correction term and \textit{t} is the time used to collect the experimental data block~\cite{Lim_concise_2014}.
The key rates displayed in~\ref{fig:keyRates}c) show an experimental secure key rate of 1.26 megabits per second at an equivalent distance of 40~km (assuming optical fibre with a 0.2~dB/km loss) using an attenuator and 246 kilobits per second in real fibre of length 75~km.
A positive secure key rate could be achieved up to 250~km in the asymptotic limit and up to 180~km with the finite key-size analysis of Eq~\ref{eq:keyRate} for just 40 minutes of key time.

To compare between polar BB84 and phase-encoded BB84, we looked at the secure key rate in the asymptotic limit using the experimental parameters obtained in the ZX and the YX bases respectively.
The transmission basis probabilities are renormalised to allow for a fair comparison.
The key rate is improved by 1.60 times when using the ZX bases compared to the YX bases.
Also, phase-encoded BB84 is able to reach an attenuation of 48.5~dB with a positive secure key rate, whereas polar BB84 can reach slightly further, at 50.1~dB.

\section{Discussion}

Our implementation produced six states in order to show the versatility of the source, however only used four for the QKD implementations.
Six-state QKD is possible and has a slightly higher tolerance to noise than its four-state counterpart, leading to the ability to share secure keys at slightly longer distances~\cite{Scarani_security_2009}.
The main drawback is found in the receiver.
Polar BB84 can be implemented with three SPDs (phase-encoded BB84 would require two SPDs in an active receiver and four SPDs in a passive receiver).
Six-state QKD, on the other hand, requires an extra AMZI and two extra SPDs if it is to remain passive, or a high speed PM in the AMZI to choose the basis in an active implementation.
Both of these options add significant complexity when compared to their meager increase in secure key distance.
Indeed, four-state QKD could be carried out for a longer time-period to reduce statistical fluctuations and increase the achievable distance.
Reference frame independent-QKD~\cite{Laing_reference-frame-independent_2010} is another protocol that requires three bases, allowing two bases to drift in time while the other stays constant.
This basis drift is a problem for polarisation-encoded systems in real fibre, however is not an issue for phase encoded systems like the one demonstrated here because the signal travels along the fibre with the phase reference.
\textcolor{black}{Multi-protocol transmitters have also been demonstrated in~\cite{Sibson_chip-based_2017, Korzh_high-speed_2013}, although these have the drawbacks of being complex and not offering phase randomisation, respectively.}

As well as the aforementioned benefit of requiring one fewer SPD for polar BB84 compared to phase-encoded BB84, the key rates are also improved by a factor of 1.60 times.
This is made possible by the reduced QBER of the signal basis from 2.8\% to 0.2\%, which reduces the bits lost to error correction, hence improving the secure key rate.

Direct preparation of decoy states can be realised by driving the pulse preparation laser at different levels above the lasing threshold to reach different intensities.
This is ideal because no external hardware, for example an intensity modulator, is required.
This would also be useful for classical communications, increasing the number of bits encoded per symbol~\cite{Noguchi_modulation_1986}.
We have achieved intensity modulation using this method.
This creates a patterning effect for the decoy states, however, where the intensity of a pulse is correlated with the intensity of the preceding pulse, opening the door to side-channel attacks~\cite{Yoshino_quantum_2018}.
To avoid this security loophole, the QKD decoy states are instead produced using a \textcolor{black}{two-level Sagnac-based IM~\cite{Roberts_patterning-effect-free_2018}}.

The transmitter also shows promise to be useful in classical communications.
The patterning effects that proved prohibitive for QKD when directly producing multiple intensity states are not a major concern here, it will just add a slight degradation to the distinguishability between states.
The direct modulation means that the system is not reliant on multiple external modulators, making it cheaper, less complex and also easier to integrate with other components.
The optical injection locking ensures that all pulses have the same wavelength.
This removes a side channel for QKD, but also means the system has low chirp, reducing the inter-symbol interference caused by dispersion effects.
In this paper we have shown the accurate production of four phase states, however this can easily be increased by using more phase-preparation levels.
Different intensities can be produced directly, and also a vacuum state can be produced, further increasing the amount of information encoded per symbol.

\section{Conclusion}

In this paper, we have demonstrated a transmitter capable of performing all weak coherent pulse-based QKD protocols, performing phase and intensity encoding simultaneously.
With this system, we have demonstrated the decoy-state polar BB84 protocol and the decoy-state phase-encoded BB84 protocol in a single experiment, preparing the six states in three different bases required by the simultaneous execution of these two protocols from a single transmitter.
In both bases we found a secure key rate in the order of 1 megabit per second at 8~dB attenuation, with the decoy-state polar BB84 protocol providing a 1.6 times larger secure key rate on average and a slightly higher tolerance to losses.

The ability to adapt to different protocols with simple software changes makes the transmitter more robust in network scenarios where all the users could potentially have different receivers.
Alongside this, the system is more simple than many other transmitters that are dedicated to a single protocol, which is appealing for real world implementations.
Also the relatively few components ensure it has a good power efficiency and make it ideal for on-chip implementations.
The versatility, low power consumption and stability of this transmitter make it the natural choice for use in metropolitan multi-user quantum networks.

\section*{Acknowledgments}
G. L. R gratefully acknowledges financial support from the EPSRC CDT in Integrated Photonic and Electronics Systems, Toshiba Research Europe Limited and an industrial fellowship with The Royal Commission for the Exhibition of 1851.
This work has been supported by funding through the EPSRC Quantum Communications Hub EP/M013472/1.

\bibliography{FullBib}

\end{document}